\documentstyle[aps,prl,epsf,twocolumn]{revtex}
\begin{document}
\def\bea{\begin{eqnarray}}
\def\eea{\end{eqnarray}}
\def\a{\alpha}
\def\d{\delta}
\def\p{\partial}
\def\nn{\nonumber}
\def\r{\rho}
\def\rv{\bar{r}}
\def\la{\langle}
\def\ra{\rangle}
\def\e{\epsilon}
\def\o{\omega}
\def\n{\eta}
\def\g{\gamma}
\def\break#1{\pagebreak \vspace*{#1}}
\def\f{\frac}
\draft
\twocolumn[\hsize\textwidth\columnwidth\hsize\csname
@twocolumnfalse\endcsname
\title{The Writhe Distribution Of Stretched Polymers}
\author{Supurna Sinha$^{}$ \cite{SUP}}
\address{Harish Chandra Research Institute,\\
Chhatnag Road, Jhunsi, Allahabad 211 019, India\\
and\\
Raman Research Institute,\\
Bangalore 560080, India.\\}
\maketitle
\widetext
\begin{abstract}
Motivated by experiments in which
single DNA molecules are stretched and twisted
we consider a perturbative approach around very high forces, where we 
determine the writhe distribution in 
a simple, analytically tractable model. Our results 
are in agreement 
with recent simulations and experiments. 
\end{abstract}

]
\narrowtext

In recent years, there has been much interest in the area of statistical
mechanics of semiflexible polymers. These studies have been motivated by 
experiments\cite{bust} on biopolymers in which single molecules are 
stretched and twisted to measure elastic properties. These 
experiments are designed to understand the role of semiflexible polymer
elasticity in, for instance, the packaging of these polymers in a cell 
nucleus. Twist elasticity plays an 
important role in several biological functions. 
The first step in packaging DNA in a cell nucleus a few microns 
across involves DNA-histone 
association which makes use of supercoiling in an
essential way.  
The process of DNA
transcription can generate and be regulated by supercoiling\cite{Strick}.
Here we focus on a particular 
class of experiments which probes the twist elasticity of DNA.  

In the experiments of Strick et al\cite{Strick}
the ends of a single molecule of double stranded DNA are attached to
a glass plate and a magnetic bead.
Magnetic fields are used to rotate
the bead and magnetic field gradients to apply forces on the bead.
By such techniques the molecule is stretched and twisted and the extension 
of the molecule is monitored by the location of the
bead. One thus gets the extension
of the molecule as a result of the applied twist and force. 

The simplest theoretical model used to interpret the 
experiments makes use of the fact that the molecule is under such high
tension that it is essentially straight. 
We call this limit the {\it paraxial limit} of the elasticity of a semiflexible
polymer keeping the optical analogy in mind \cite{BM,mmg}.
In this situation, the molecule, being straight and taut, cannot 
intersect itself and so one does not expect 
self-avoidance effects to be important. In computing the 
partition function one simply sums over all
configurations without regard to self-avoidance. 
This model
is instructive because it is analytically tractable and 
enables us for {\it the first time} to derive a simple
explicit analytic expression for the writhe distribution. 
This is the central result of our analysis. 

Before discussing the schematics of the derivation of the writhe
distribution let us define a few pertinent quantities: 
link, twist and writhe. For an open polymer of the kind that is 
used in the twist-stretch experiments one can define link ($Lk$),
an arbitrary real number, as the externally imposed twist 
given by $2\pi n$ where $n$ is the number of applied
turns on the bead. Twist ($Tw$) corresponds to 
the integrated rotation of the 
polymer around its backbone and writhe ($Wr$) pertains to the 
twist of the polymer backbone and is captured by the rotation of 
the tangent vector. In computing the writhe distribution  
we make use of the fact that link $Lk$ 
is related to twist and writhe via $Lk = Tw +Wr$ \cite{twist,Cal,fuller}.
It is convenient to go to the conjugate space and work with the variable
$B$, the generator of link $Lk$. In this space the partition 
function
$Z(B,f)$ neatly factors into writhe
$Z_W(B,f)$ and twist $Z_T(B,f)$ \cite{Bouchiat}. The distribution 
${{\tilde Z}(Lk,f)}$
is a convolution of the writhe distribution and the twist distribution.
We compute the link distribution in the $\alpha=L_{BP}/L_{TP}$ 
(the ratio of the 
bend and the twist persistence lengths)
$\rightarrow 0$ limit. In this limit the writhe distribution coincides 
with the link distribution.
This is simply because in the $\alpha \rightarrow 0$ limit it is very 
expensive to twist the polymer around a straight backbone and the polymer
goes into a bending mode resulting in a twisting of the polymer backbone
\cite{twist}.
From the  writhe distribution 
computed in this manner, one can recover the link distribution  
by convolving it with the distribution of ``dynamical twist''.

Our starting point is the Worm Like Chain ({\it WLC}) model 
in which the polymer is modelled as
a framed space curve ${\cal
C}=\{\vec x(s),{\hat e}_i(s)\}, i=1,2,3,$ where $0\le s\le L$ is the
arclength parameter along the curve. 
The unit tangent
vector ${\hat e}_3= {d{\vec x}/ds} $ to the curve describes the
bending of the polymer while the twisting is captured by a
unit vector
${\hat e}_1$ normal to ${\hat e}_3$.
In keeping with the optics analogy \cite{BM,mmg},
we refer to ${\hat e}_1$ as the
polarisation vector.
${\hat e}_2$ is then fixed by
${\hat
e}_2={\hat e}_3
\times {\hat e}_1$ to complete the right handed moving frame
${\hat e}_i(s), i = 1,2,3 $.
The energy ${\cal E}[{\cal C}]$ of a configuration of the polymer is a 
sum of contributions coming from its bending and twisting modes. 

In the presence of large forces $|\vec{F}| \rightarrow \infty$, the
molecule is stretched taut and there is an energy barrier for the 
polymer to pass through itself. 
In fact, the molecule is constrained to lie in
nearly a straight line between its ends with small deviations.
As mentioned earlier, in this regime the polymer 
being essentially straight cannot cross
itself and thus self-avoidance effects present in a real polymer
are automatically taken into consideration.
Under these conditions the tangent vector only makes small deviations from
the $\hat{z}$ direction. We can approximate the sphere 
of directions\cite{twist}
by a tangent plane
at the north pole of the sphere. 
We call this limiting model of
the WLC the paraxial worm like chain or the PWLC model.
  
In this straight taut limit
the polymer Hamiltonian\cite{twist,Bouchiat,wilhelm,Nelson} reduces to :
$$H_{PWLC}=\frac{p_{\theta}^2}{2}+\frac{(p_{\phi}-A_{\phi})^2}{2{\theta}^2}
+\frac{B^2\alpha}{2}-f(1-\frac{{\theta}^2}{2})$$
$$=H_P -f +\frac{B^2\alpha}{2}$$
where $H_P$ is the Hamiltonian of interest in the paraxial limit after
we take out a constant piece. $\alpha$ is the ratio of the 
bend persistence length $L_{BP}$ and the twist persistence length 
$L_{TP}$. The constant $B$ corresponds
to the conserved momentum conjugate to the Euler angle  
$\psi$.
$f=FL_{BP}/k_BT$ where $F$ is the 
stretching force and $k_BT$ is the thermal energy.
The `vector potential' 
$A_{\phi}=B\frac{\theta^2}{2}$. 
Thus the PWLC maps on to the problem of a particle moving on 
a plane in the presence of a magnetic
field $B$ and an oscillator confining potential which arises
from making a small $\theta$ expansion for
the stretching force $-f \cos{\theta}$
($-f\cos{\theta}\approx -f(1-\frac{{\theta}^2}{2})= 
-f+f\frac{{\theta}^2}{2}$)\cite{Bouchiat,Nelson}.
Notice that, in contrast to the regime of low tension \cite{twist}, in 
this high tension regime the polymer cannot 
release an imposed twist by passing 
through itself because of the condition of suppression of configurations
in which the polymer folds back onto itself. This implies that
in contrast to the WLC model\cite{twist}, in the PWLC model 
the free energy, torque-twist relation and other related distributions
are {\it not} periodic functions of the imposed twist.

Introducing Cartesian coordinates $\xi_{1}=\theta \cos{\phi}$ and
$\xi_{2}=\theta \sin{\phi}$
on the tangent plane $R^2$ at the north pole
of the sphere of directions
one can express the small $\theta$ Hamiltonian $H_P$ as
follows:
\begin{equation}
H_P=\frac{1}{2} {(p_{\xi_{1}}-A_{\xi_{1}})}^2
+\frac{1}{2} {(p_{\xi_{2}}-
A_{\xi_{2}})}^2+\frac{f}{2}({\xi_{1}}^2+{\xi_{2}}^2)
\label{hampwlcp}
\end{equation}
where $A_{\xi_{1}}=-B{\xi_{2}}/2,A_{\xi_{2}}=B{\xi_{1}}/2$.
The corresponding partition function $Z$ can be written
in terms of the eigenvalues $E_n$ and eigenfunctions $\{u_n\}$
of $H_{PWLC}$ as follows:
$$Z(B,f,\vec{\xi_{0}},\vec{\xi_{L}})=\sum_{n}{e^{-\beta E_n(B,f)}
u_n^{*}(\vec{\xi_{0}})u_n(\vec{\xi_{L}})}$$
where $\vec{\xi_{0}}=(\xi_{1}{(0)},\xi_{2}{(0)})$
and $\vec{\xi_{L}}=(\xi_{1}{(L)},\xi_{2}{(L)})$ are the initial and final
tangent vectors at the two ends of the polymer.
In order to simplify our analysis further we confine ourselves to the
limit
of very long polymers. Many of the experiments involving biopolymers such 
as
DNA explore this limit of very long polymers which is also theoretically
more tractable. 
For long polymers ($\beta=\frac{L}{L_{BP}}\rightarrow \infty$)
only the lowest eigenvalue 
$E_0(B,f)={\sqrt{f+{{B}^{2}}/4}}-f+\frac{B^2\alpha}{2}$ 
\cite{Nelson,fock}
of $H_{PWLC}$
dominates the expression for the partition function.
Thus the partition function can be written as:
\begin{equation}
Z(B,f)_{\beta \rightarrow \infty}= e^{-\beta E_0(B,f)}
\label{lbeta1}
\end{equation}

{\bf Writhe Distribution:}

Let us consider the link distribution $\tilde{Z}(Lk,f)$:
\begin{equation}
\tilde{Z}(Lk,f)= \int{e^{-\beta E_0(B,f)-iB Lk} dB}
= \int{e^{i\phi(B,Lk,f)} dB}
\label{gener}
\end{equation}
where the phase $\phi(B,Lk,f)$ is given by
$\phi(B,Lk,f) = {i\beta E_0(B,f)-\beta B\frac{Lk}{\beta}}$.
Since we are working in the limit of long polymers we can compute this
partition function using the stationary phase or saddle point 
method\cite{Bouchiat} where
only the stationary value $\phi(B_{st},Lk,f)$ of the phase dominates.
$\phi(B_{st},Lk,f)$ is the {\it central quantity of interest} from 
which all
the
relevant elastic properties of the taut polymer can be derived.
A similar
perturbative analysis was done by Moroz and Nelson\cite{Nelson}, who in 
fact carry the
analysis out to higher orders in perturbation theory. What is new in our 
treatment is an {\it explicit analytical expression} for the writhe 
distribution in the straight taut limit.

Here we outline the derivation for the writhe distribution.
The condition for stationarity ($\frac{\partial{\phi}}{\partial{B}}=0$)
satisfied by the stationary value $B_{st}$
of $B$ is:
\begin{equation}
\frac{(i\tilde{Lk}-\alpha{B_{st}})}{B_{st}/4}= 
\frac{1}{\sqrt{f+B_{st}^2/4}}.
\label{cfinite}
\end{equation}
where $\tilde{Lk}=\frac{Lk}{\beta}$.
We restrict to the case of $\alpha=0$.
In this case the equation simplifies and we get the following
stationary value of $B_{st}$:
$$B_{st} = \pm i\frac{4 
\sqrt{f}{\tilde{Lk}}}{\sqrt{[1+4{\tilde{Lk}}^2]}}$$
Of the two roots, only the positive root is the physically 
relevant one consistent with the saddle point approximation. 
Setting $iB_{st}=\tau$, where $\tau$ has the interpretation of torque, we 
get 
the following Torque-Link relation:
\begin{equation}
\tau = \frac{4
\sqrt{f}{\tilde{Lk}}}{\sqrt{[1+4{\tilde{Lk}}^2]}}
\label{torque}
\end{equation}
Notice that for small ${\tilde{Lk}}$ we get a linear Torque-Link 
$(\tau-{\tilde{Lk}})$ relation which goes over to a Torque-Link relation 
independent of ${\tilde{Lk}}$ in the limit of large ${\tilde{Lk}}$.
This is consistent with recent experiments\cite{bust2} and
numerically generated plots\cite{Bouchiat}.
Inserting the
expression for the stationary value $B_{st}$ of $B$ into the partition 
function
$\tilde{Z}(Lk,f)$ we get the pertinent writhe distribution $P(W,f)$.
To compute this distribution we have made use of the fact that for 
$\alpha \rightarrow 0$, twist is extremely expensive and the applied twist
goes completely into the bending mode. Thus, the link distribution in 
this limit corresponds to the distribution $P(W,f)$ of writhe.  
Given the writhe distribution $P(W,f)$ obtained in this manner 
the link distribution  $P(Lk,f)$ can be constructed   
for {\it all} values of $\alpha$ by convolving it with the twist 
distribution. In the generating function space one simply needs to 
multiply the writhe partition function $Z_{W}(B,f)$ by a simple 
Gaussian factor $ Z_{T}(B,f)= e^\frac{-\alpha B^2}{2}$
pertaining to the pure twist distribution at finite $\alpha$. 

The analytic form of the scaled distribution 
$P(W)= \frac{P(W,f)}{P(0,f)}$
of writhe and stretch in 
the high
tension regime (See Fig. $1$ ) is:
\begin{eqnarray}
P(W)=\exp[-{\beta\sqrt{f}\{\sqrt{1+4W^2}-1\}}]
\label{writhe}
\end{eqnarray}
Here $f=\frac{FL_{BP}}{k_BT}$ with $F$ the applied stretching force. 
This analytic form of the writhe distribution is the central result 
of this paper. Plots of this distribution are displayed in Fig. $1$. 
The form reduces to a Gaussian form 
($P(W)\approx\exp[-{2{\beta}{\sqrt{f}} W^2}]$) for small 
values 
of 
writhe $W$
and goes over to 
$P(W)\approx\exp[-{2{\beta}{\sqrt{f}}|W|}]$ for very
large values of writhe $W$.
The writhe distribution (Eq. (\ref{writhe})) represented 
in Fig. $1$ has all the 
expected features- it peaks near smaller values and dies off
for larger values of writhe. Writhe gets suppressed with increasing 
strength
of the stretching force. These qualitative features are in agreement 
with recent simulations\cite{volo} of writhe as a 
function of the stretching force. The explicit expression for the writhe
distribution presented here is exact in the high tension limit and we 
expect quantitative agreement between the predicted distribution and 
future experiments probing the writhe distribution in this regime. 

To summarize, we have, for the first time obtained an explicit 
analytical expression for 
the writhe distribution of a semiflexible polymer in the high tension 
regime. 
The expression for the writhe distribution 
is simple and transparent
and the qualitative features agree well with  
available computer simulations \cite{volo}. 
For very large forces a DNA molecule undergoes force 
induced denaturation\cite{Strick}
and therefore the distribution predicted here may not be directly 
applicable to DNA experiments at very high tension. However, one
can test the predictions against experiments with other semiflexible polymers.
We therefore, expect this work to generate interest amongst experimentalists 
to measure the writhe distribution of stretched polymers.
The distribution computed here is also relevant to depolarized light
scattering in turbid media in the 
limit of small angle scattering\cite{BM,mmg,maggs}. 
We also have an explicit analytic form for the 
Torque-Twist relation which is 
in agreement with recent experimental data\cite{bust2}. 
In this model the mean-squared writhe 
fluctuation which corresponds to the second derivative of the conjugate
distribution $\tilde{Z}(Lk,f)$, diverges at $\tau=2\sqrt{f}$, the point
at which a `buckling instability' sets in for the polymer.  
More explicitly, the divergence 
of the second moment of the writhe distribution at the buckling 
instability point $\tau=2\sqrt{f}$
has the following form which can be tested 
against future experiments:
$$<W^2>=\frac{f}{4[f-\frac{\tau^2}{4}]^{3/2}}.$$
The second moment of the distribution is consistent with
earlier predictions for the mean squared writhing angle for 
long tense molecules\cite{maggs2}. At the buckling instability
there is a divergence of the writhe fluctuations. Since such a 
divergence makes the polymer backbone fluctuate violently one
expects it to lead to a corresponding divergence in the mean 
squared extensional fluctuations $<\xi^2>$. 
This has, in fact been probed in 
some recent experiments\cite{benchar}. 
In the `paraxial' limit we find that $\frac{<\xi^2>}{<W^2>}=\frac{1}{f}$.  

The `paraxial' approximation breaks down for large values of the 
applied twist 
(more precisely for $\tau>2\sqrt{f}$ [See Eq. {\ref{cfinite}}] ) in which 
case the polymer
explores configurations which deviate considerably 
from the straight taut limit. 
In future we would like to explore the low tension nonlinear 
regime for the writhe
distribution where phenomena like plectoneme formation would
play an important role and 
nontrivial self-avoidance effects\cite{BM,maggs2,kamien,kamien2} need to 
be
taken into consideration. The present work will 
provide a limiting check on calculations done in the nonlinear regime.  
As we mentioned earlier, the writhe distribution has important 
implications in the context of transcription and gene regulation.
Therefore, a complete understanding of writhing of a biopolymer backbone
and its stabilization is of relevance to current research.

\vbox{
\begin{figure}
\epsfxsize=6.0cm
\epsfysize=6.0cm
\epsffile{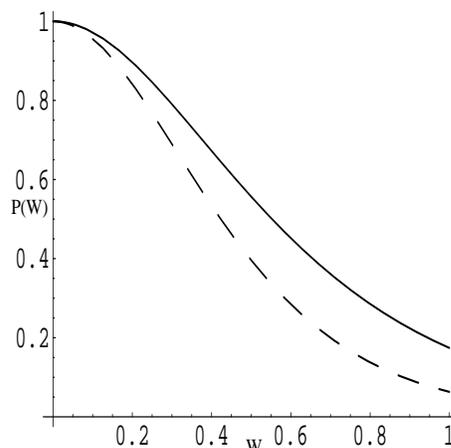}
\caption{The Writhe distribution $P(W)$ for $f=2$ and 
$f=5$(dashed 
curve) for $\frac{L}{L_{BP}}=10.$ }
\end{figure}}

{\it Acknowledgements:}
It is a pleasure to thank G. Charvin and 
D. Bensimon for making some of their data
available before publication 
and A. Maggs, D. Bensimon, Y. Rabin, D. Dhar, 
M. Randeira, J. Samuel and A. Dhar for discussions.

\end{document}